\documentstyle[epsfig]{mn2e}

\def\be{\begin{equation}}
\def\ee{\end{equation}}
\def\bea{\begin{eqnarray}}
\def\eea{\end{eqnarray}}
\def\etal{et al. }

\title[]{Cosmic ray protons in the energy range $10^{16}-10^{18.5}$ eV:
stochastic gyroresonant acceleration in hypernova shocks?}
\author[]{Yi-Zhong Fan$^{1,2}$\thanks{E-mail: yzfan@pmo.ac.cn}\\
$^1${\sl Niels Bohr International Academy, Niels Bohr Institute,
Copenhagen University, Blegdamsvej 17, DK-2100 Copenhagen,
          Denmark }\\
$^2${\sl Purple Mountain Observatory, Chinese Academy of
Sciences, Nanjing 210008, China}\\
}
\date{Accepted 23 June 2008;
Received 23 June 2008; in original form 5 December 2007}

\begin{document}

\maketitle
\begin{abstract}
The hypernovae (HNe) associated with Gamma-ray Bursts (GRBs) may
have a fairly steep energy-velocity distribution, i.e., $E(\geq
\beta)\propto \beta^{-q}$ for $q<2$ and $\beta\geq \beta_o$, where
$\beta$ is the velocity of the material and $\beta_o \sim 0.1$ is
the velocity of the slowest ejecta of the HN explosion, both in
units of the speed of light $(c)$. The cosmic ray protons above the
second knee but below the ankle may be accelerated by the HN shocks
in the velocity range of $\beta \sim \beta_o - 4\beta_o$. When
$\beta \leq 4\beta_o$, the radius of the shock front to the central
engine is very large and the medium decelerating the HN outflow is
very likely to be homogeneous. With this argument, we show that for
$q\sim 1.7$, as inferred from the optical modelling of SN 2003lw,
the stochastic gyroresonant acceleration model can account for the
spectrum change of high energy protons around the second knee. The
self-magnetized shock acceleration model, however, yields a too much
steep spectrum that is inconsistent with the observation unless, the
medium surrounding the HN is a free wind holding up to a
(unrealistic large) radius $\sim 1-10~{\rm kpc}$ or alternatively
the particle acceleration mainly occurs in a narrow ``dense" shell
that terminates the free wind at a radius $\sim 10^{19}$ cm.
\end{abstract}

\begin{keywords}
acceleration of particles $-$ cosmic rays $-$ Gamma Rays: bursts $-$
supernovae: general $-$ supernova remnants
\end{keywords}

\section{Introduction}
One of the most widely suggested sources of cosmic rays (CRs) is the
supernova (SN) remnants (see Hillas 2005 for a review). Since 2001,
more and more researchers have noted that hypernovae (HNe), in
particular those associated with gamma-ray bursts (GRBs), may play
an important role in interpreting the CR spectrum above its first
knee\footnote{The second knee and the ankle in the CR spectrum are
at $\sim 3\times 10^{17}$ eV and $\sim 3\times 10^{18}$ eV,
respectively.}, i.e., $\sim 3\times 10^{15}$ eV (Dermer 2001a,
2001b; Erlykin, Wibig and Wolfendale 2001; Sveshnikova 2003; Wick,
Dermer and Atoyan 2004; Wang et al. 2007; Budnik et al. 2008). This
is reasonable since the average velocity and the total kinetic
energy of HN outflows are much larger than that of normal SNe. So is
the energy of accelerated particles. A reliable interpretation of CR
spectrum up to $\sim 10^{18}$ eV thus should take into account the
different energies and types among SNe (Sveshnikova 2003).

How to accelerate protons up to an energy $\sim10^{19}$ eV in the HN
blast waves? Dermer (2001b) suggested that the gyroresonant
stochastic acceleration might play such a role (see Fig.10 of Dermer
2001a for a quantitative plot). Other authors (Wang et al. 2007;
Budnik et al. 2008; Erlykin et al. 2001) considered the
self-magnetized acceleration model put forward by Bell \& Lucek
(2001), in which the magnetic field of the upstream has been
significantly amplified by CRs. Considering the energy distribution
of the HN outflow\footnote{See Berezhko \& V\"olk (2004) and Ptuskin
\& Zirakashvili (2005) for the influence of the energy distribution
of normal SN outflows on the spectrum of accelerated CRs.}, Wang et
al. (2007) and Budnik et al. (2008) suggested that with the second
model the CR proton spectrum steepening around the second (first)
knee could be reproduced. In this work, we point out one potential
limit of such an interpretation and show that the gyroresonant
stochastic acceleration model does not suffer from that problem.

This paper is arranged as follows. In section 2 we discuss the
energy-velocity distribution of HN outflows and the medium profile
surrounding the HN outflows. We find that for HN outflows owning a
fairly steep energy-velocity distribution, when $\beta_o\leq \beta
\leq 4\beta_o$ that may play the main role in accelerating the CR
protons above the second knee but below the ankle (see the
discussions below eq.(\ref{eq:Wang}) and eq.(\ref{eq:3})), the
radius of the shock front to the central engine is very large and
the medium decelerating the HN outflow is likely to be homogeneous.
In section 3 we calculate the change of the CR spectrum around the
second knee which is caused by the energy-velocity distribution of
HN outflows, and compare the results with the CR spectrum
observation so as to constrain the models. In Section 4, we
summarize our results with some discussions.

\section{The hypernova outflow: Energy-velocity distribution and the medium it expands into}
\subsection{Energy-velocity distribution of hypernova outflows}
HNe, especially those of which associated with GRBs/XRFs, are
distinguished for the broad lines in their spectra, indicating very
high expansion velocity of the ejecta. The modelling of optical
light curves and spectra, in principle, can reconstruct the
energy-velocity distribution of the outflows. However, no reliable
constraint can be given on the $\beta>0.3$ outflow by optical even
if that part had some optical depth since the current optical
modelling is not fully relativistic (J. S. Deng 2008, private
communication). In SN 1998bw, SN 2003dh, SN 2003lw and SN 2006aj,
strong photospheric velocity evolution (the inner the outflow, the
smaller the velocity) are evident \cite{Hjorth03,Mazzali05,sod06}.
The optical modelling of SN 2003lw gave that material moving faster
than $(0.1,~0.2)c$ were $\sim (1.4, ~0.1)M_\odot$, respectively
(Mazzali et al. 2006), implying a fairly steep initial kinetic
energy distribution $E(\geq \Gamma \beta) \propto (\Gamma\beta
)^{-1.7}$, where $\Gamma=(1-\beta^2)^{-1/2}$. But in other events,
no result has been published. Soderberg et al. (2006) constrained
the kinetic energy profile of HN outflows in a more speculative way.
They used optical spectral data to probe the slowest ejecta in
supernova explosions and employed the radio observation to trace the
fastest component of the outflow. They then took these two data
points to estimate the energy-velocity distribution. Their results
may be biased because the fast moving material identified by radio
observation might be the decelerated GRB/XRF ejecta rather than the
fastest component of the main SN explosion. If so, it is not a
continuous distribution of matter between the two data points
\cite{sod06,Xu08}.

Fairly speaking, observationally so far we do not have a reliable
estimate of the initial kinetic energy-velocity distribution of
(most) HN outflows in the velocity range of $\beta\sim 0.1-0.5$.
Theoretically the standard hydrodynamic collapse of a massive star
(Tan, Matzner \& Mckee 2001) results in a kinetic energy profile of
the SN explosion $E(\geq \Gamma \beta) \propto (\Gamma
\beta)^{-5.2}$. Such a steep function, however, is inconsistent with
the constraint from the optical data of SN 2003lw (Mazzali et al.
2006), for which a rough estimate gives $E(\geq \Gamma \beta)
\propto (\Gamma\beta )^{-1.7}$. Motivated by this fact, we {\it
assume} that all HNe associated with GRBs have a fairly steep energy
distribution, which is generally written as $E(>\Gamma \beta)
=A[\Gamma \beta/(\Gamma_o \beta_o)]^{-q}$ for $\beta<0.5$, where
$\Gamma_o=(1-\beta_o^2)^{-1/2}$. For SN1998bw/GRB980425,
SN2003dh/GRB030329, SN2003lw/GRB031203, and SN2006aj/GRB060218, {\it
optical modelling} suggests $A\sim 0.2-6\times 10^{52} {\rm erg}$
and $\Gamma_o \beta_o \sim 0.04-0.1$ \cite{sod06}. The parameter
$q$, however, is not reliably determined in most cases. For
simplicity, we approximate $E(>\Gamma \beta) =A[\Gamma
\beta/(\Gamma_o \beta_o)]^{-q}$ as $E(>\beta) =A(\beta/
\beta_o)^{-q}$ for $\Gamma \sim 1$.

In a SN explosion, the larger the velocity, the outer the ejecta.
When the fast component is decelerated by the medium, the slower
part will catch up with the decelerating shock front. As a result,
the total kinetic energy of the shocked medium increases and the
deceleration of the shock is suppressed. In the quasi-similar
evolution phase of the HN shock, the fastest component has swept
enough medium and has got decelerated. Significant part of the
initial kinetic energy of the HN material $E(\geq \beta)$ has been
used to accelerate the medium to a velocity $\sim \beta$. So when we
talk about the CR acceleration in the blast wave, the $E(\geq
\beta)$ mentioned there actually represents the total kinetic energy
of the shocked medium moving with a velocity $\beta$. For a medium
taking the profile $n\propto R^{-k}~(0\leq k \leq 3)$, the rest mass
swept by the HN blast wave is $M_{\rm med}=\int^{R}_0 4\pi n m_{\rm
p} R^2 dR\propto R^{3-k}$. For $\beta>\beta_o$, conservation of
energy gives $E(\geq \beta) \approx M_{\rm med}\beta^2/2$, i.e.,
\begin{equation}
\beta^{-(q+2)} \propto R^{3-k}.\label{eq:energy_conser}
\end{equation}
With the relation that $R \sim \beta t$, the dynamics of the HN
outflow is described by (for $\beta>\beta_o$)
\begin{equation}
\beta \propto t^{-{3-k\over5+q-k}}. \label{eq:beta}
\end{equation}
In the next section we'll show how $q$ and $k$ influence the CR
spectrum.

\subsection{The medium the hypernova outflow expands into}
As shown in Eq. (\ref{eq:beta}), the dynamics of a HN shock depends
on the medium profile sensitively. Here we review the medium
profiles of all four GRB-HN events, based on the GRB and/or HN
afterglow modeling. For GRB 980425, the medium is Wind like (i.e.,
$k \sim 2$). The afterglow modeling favors an unusual small $A_*
\sim 0.01-0.04$ \cite{LC99,W04}, where $A_*\equiv
(\dot{M}/10^{-5}M_\odot~\rm yr^{-1})(v_w/10^{8}~\rm
cm~s^{-1})^{-1}$, $\dot{M}$ is the mass loss rate of the progenitor,
and ${\rm v_w}$ is the velocity of the stellar wind. For GRB 030329,
the circumburst medium is found to be homogeneous (i.e., $k \sim
0$), as shown in many independent investigations
\cite{Frail05,Pih07,van07,Xue08}. For GRB 031203, after modeling the
radio data, Soderberg et al. (2004) got a constant \footnote{There
is an additional evidence for such a conclusion. With an electron
energy distribution index $p\sim 2.6$ inferred from the radio data,
the X-ray spectrum $F_\nu \propto \nu^{-0.8}$ \cite{sod04a} suggests
that the X-ray emission is below the cooling frequency of the
forward shock. For a free-wind medium, the late-time ($t>1$ day)
X-ray light curve should drop with the time as $t^{-1.7}$, deviating
from the detected $t^{-1}$ decline (e.g., Ramirez-ruiz et al. 2005)
significantly. If the medium is ISM-like, the expected late-time
decline is $t^{-1.2}$, consistent with the data. In this scenario,
the early time ($t<1$ day) X-ray flattening, as those well detected
in {\it Swift} GRBs, can be attributed to an energy injection from
the central engine.} $n\sim 0.6~{\rm cm^{-3}}$. For GRB 060218, the
high quality radio data supports the homogeneous medium model with
$n\sim 100~{\rm cm^{-3}}$ \cite{FPX06,sod06}. As such, we have no
compelling evidence for a Wind-like medium surrounding most GRBs,
even for those associated with HNe. The physical reason is not
clear, yet. A post-common envelope binary merger model (e.g., Fryer,
Rockefeller \& Young 2006) or a fast motion of the Wolf-Rayet star
relative to the ISM (van Marle et al. 2006) may be able to solve
this puzzle.

Actually, a free wind medium, supposed to surround the progenitor,
is unlikely to be able to keep such a profile up to the radius
\begin{equation}
R_{\rm dec}(\beta_o) \sim 3\times 10^{22}~({M_{\rm ej}\over
10M_\odot})A_{*,-1}^{-1}~{\rm cm},
\end{equation}
where $M_{\rm ej}$ is the rest mass of the GRB-associated HN ejecta,
and $R_{\rm dec}$ is the deceleration radius. This is because during
their evolution, massive stars lose a major fraction of their mass
in the form of a stellar wind. The interaction between this stellar
wind and the surrounding interstellar medium creates a circumstellar
bubble (e.g., Wijers 2001; Ramirez-Ruiz et al. 2001; Dai \& Wu 2003;
Chevalier et al. 2004; van Marle et al. 2006). The analytical
calculation suggests that the free wind of a Wolf-Rayet star usually
terminates at \cite{C07}
\begin{equation}
 R_{\rm t} = 5.7 \times 10^{18} ~
({v_w \over 10^3~{\rm km ~s^{-1}}}) ({p/k\over 10^5 {\rm cm^3 ~
K}})^{-1/2}A_{*,-1}^{1/2}~{\rm cm},
\end{equation}
where $p$ is the pressure in the shocked wind and $k$ is the
Boltzmann constant. This is confirmed by observations of Wolf-Rayet
nebulae, such as NGC 6888 and RCW 58, which also have radii of the
order of a few pc (Gruendl et al. 2000). Here we take the numerical
example given in Figure 1 of Chevalier et al. (2004) to show that
the HN outflow is mainly decelerated in the ISM-like medium region.
In their numerical example, $n\sim 0.5R_{18}^{-2}~{\rm cm^{-3}}$ for
$R_{18}<1.2$. The total mass of the free wind medium is thus $\sim
7\times 10^{-3}~M_\odot \ll M_{\rm ej}$. Here and throughout this
work, the convention $Q_{\rm x}=Q/10^{\rm x}$ has been adopted in
cgs units.

We therefore conclude that the medium is most likely to be ISM-like
at the radius where the HN outflow has been decelerated to $\beta <
4\beta_o$. This can be also understood as follows. One can infer
from eq.(\ref{eq:energy_conser}) that for $k=2$ the outflow
component with a $\beta>\beta_o$ will decelerate at a radius $R_{\rm
dec} (\beta)\approx (\beta/\beta_o)^{-(2+q)}R_{\rm dec}(\beta_o)$.
So for $\beta \sim 4\beta_o$ and $q\sim 2$,
\[R_{\rm dec}(4\beta_o) \sim 4\times 10^{-3}R_{\rm dec}(\beta_o) \sim R_{\rm t}.
\]

\section{Spectrum of cosmic ray protons: observation and interpretation}
\subsection{A new CR proton component in the energy range of $10^{16}-10^{18.5}$ eV}
The spectrum of protons steepens suddenly at the first knee by a
factor of \[\Delta \gamma {\rm (I)} \sim -2.1.\] In view that the
spectra of heavier particles would steepen at higher energies, the
likely interpretation of the steepening of all CRs at the first knee
is the sudden decline of the light particles such as H and He (see
Hillas 2005; H\"orandel 2008 and the references therein).

The proton CR spectrum before and after the second knee, after
subtracting the modeled ``galactic" component, can be roughly
estimated as (Ulrich et al. 2004; Antoni et al. 2005; Hillas 2005;
H\"orandel 2008)
\begin{equation}
{dN_{\rm CR} \over dE_{\rm CR}} \propto
\left\{%
\begin{array}{ll}
   E_{\rm CR}^{-2.4} & {\rm for~} 0.1<E_{\rm CR,17}<3, \\
    E_{\rm CR}^{-3.3} & {\rm for~} 3 <E_{\rm CR,17}<30,\\
\end{array}%
\right. \label{eq:5}
\end{equation}
which indicates the factor of spectral steepening is $\Delta
\gamma{\rm (II)} \sim -0.9$. Above the ankle, the CR spectrum
changes to $E_{\rm CR}^{-2.7}$, so the factor of flattening is
$\Delta \gamma{\rm (III)} \sim 0.6$.

The interpretations of spectral changes at the second knee and at
the ankle are much less clear. Hillas (2005) interpreted them as a
result of an extragalactic component with a spectrum $\propto E_{\rm
CR}^{-2.3}$ suffering losses by the interaction between cosmological
microwave background radiation and starlight. In this work, we
consider the detected spectral change around the second knee is due
to the energy-velocity distribution of HN outflows.

\subsection{Theoretical interpretation}
\subsubsection{Self-magnetized shock acceleration model}\label{sec:SMSAM}
In this model, the magnetic field of the upstream is assumed to be
amplified significantly by the CRs themselves (e.g., Bell \& Lucek
2001).

With Eq. (\ref{eq:beta}), we have the radius of the forward shock
front as $R \propto t^{{2+q\over5+q-k}}$. The maximum energy
accelerated by the forward shock can be estimated by
\cite{BL01,BV04,PZ05}
\begin{equation}
E_{\rm max}(k,q)\sim Z \beta e B R \propto t^{k-4+q(2-k)/2\over
5+q-k} \propto \beta^{4-k-q(2-k)/2 \over 3-k},
\end{equation}
where $B \propto \beta R^{-k/2}$ is the magnetic field in the upstream of the shock,
which is on the same order of that of the shocked medium.

In the ISM case (i.e., $k=0$), we have
\begin{equation}
E_{\rm
max}(0,q)\propto \beta^{(4-q)/3},
\end{equation}
while in the WIND case (i.e., $k=2$),
\begin{equation}
E_{\rm max}(2,q)\propto \beta^{2}. \label{eq:Wang}
\end{equation}

Here we do not present the numerical coefficient of $E_{\rm
max}(k,q)$ because Wang et al. (2007) and Budnik et al. (2008) have
already shown that for typical parameters, a $\beta\sim \beta_o\sim
0.1$ is high enough to accelerate protons up to $\sim 10^{17}$ eV
regardless of $k$. In a stellar wind medium, the HN shock front with
a $\beta\sim 4\beta_0$ can accelerate protons up to $\sim 3\times
10^{18}$ eV. Therefore the CR protons above the second knee but
below the ankle are mainly accelerated by the HN shock in the
velocity range of $\sim \beta_o-4\beta_o$.

To get an estimate of the spectrum of the accelerated particle,
following Berezhko \& V\"olk (2004) and Ptuskin \& Zirakashvili
(2005) we assume: (1) The particles with an energy $E_{\rm max}$
escape the
 shock immediately; (2) The total energy of the accelerated
 particles at an energy $E_{\rm CR}=E_{\rm max}(\beta)$
is proportional to $E(\geq \beta)$. In view of the relations $E(\geq
\beta)\propto [E_{\rm max}(\beta)]^{-3q/(4-q)}$ for $k=0$ and
$E(\geq \beta)\propto [E_{\rm max}(\beta)]^{-q/2}$ for $k=2$, we
have
\begin{equation}
{dN \over dE_{\rm CR}} \propto
\left\{%
\begin{array}{ll}
   E_{\rm CR}^{-(2+\delta)-{3q\over
4-q}} & {\rm for~} k=0, \\
    E_{\rm CR}^{-(2+\delta)-{q\over
2}} & {\rm for~} k=2,\\
\end{array}%
\right. \label{eq:R1}
\end{equation}
where $\delta \sim 0.4$ is introduced to account for the proton
spectrum in the energy range of $10^{16}-3\times 10^{17}$ eV. As
$\beta\leq \beta_o$, $E({\geq \beta}) \propto \beta^0$ if the energy
loss of the HN shock is ignorable. The accelerated proton spectrum
should be $\propto  E_{\rm CR}^{-(2+\delta)}$. This answers why
there comes a spectrum change around the second knee if $E_{\rm
max}(\beta_o)\sim {3\times 10^{17}}$ eV.

With a $\delta = 0$, to match the detected proton spectrum
$dN/dE_{\rm CR}\propto E_{\rm CR}^{-3.3}$, one has to have $q \sim
2.6$, which is very close to that of SN 2003lw and SN 1998bw {\it
reported in Soderberg et al.}\footnote{Please see section 2.1 for
the discussion of uncertainty of the $q$ obtained in their way.}
(2006). Therefore Wang et al. (2007) concluded that the
self-magnetized shock acceleration model could account for the
spectrum data. However, a few puzzles have to be solved before
accepting this argument: (I) If $\delta = 0$, some novel effects are
needed to interpret why the proton spectrum departs from $E_{\rm
CR}^{-2}$ significantly in the $10^{16}-3\times 10^{17}$ eV range.
 The authors also need to explain why these effects, if
any, disappeared in the $3\times (10^{17}-10^{18})$ eV range. (II) A
wind profile holding to a radius $\sim 1-10$ kpc is crucial for
their argument.  If the medium is ISM-like when the outflow gets
decelerated to $\beta<0.4$, Wang et al. (2007)'s approach would
yield a spectrum
\begin{equation}
{dN \over dE_{\rm CR}} \propto E_{\rm CR}^{-5} \label{eq:puzzle}
\end{equation}
for $q\sim 2$, which is too steep to be consistent with the data. We
take this puzzle as a potential limit of their interpretation.

Let's investigate whether a specific wind-bubble can solve this
puzzle. We assume that the free wind profile is terminated at a
radius $\sim R_{\rm t} \sim 10^{19}$ cm and is followed by an
ISM-like shell. Suppose that the shell is so massive that the
deceleration of the whole HN outflow occurs at $R\sim R_{\rm t}\sim
{\rm const}$, we have $E_{\rm max}\sim Z \beta e B R \propto \beta
B$. If the shell is not dense enough to form a strong reverse shock,
i.e., the forward shock velocity decreases continually rather than
abruptly, then $B \propto \beta n^{1/2}$. As a result, we have
$E_{\rm max} \propto \beta^2 n^{1/2}$ and $dN/dE_{_{\rm CR}}\propto
E_{\rm CR}^{-(2+\delta)-{q\over 2}}$, provided that the CR protons
in the energy range of $\sim 3\times (10^{17},~10^{18})$ eV are
mainly accelerated in the shocked shell. Though such a possibility
is attractive, the request that the reverse shock does not form is
hard to satisfy. This is because at a radius $\sim R_{\rm t} \sim
10^{19}$ cm the number density of the wind medium $n_{\rm w}\sim
3\times 10^{-4}~{\rm cm^{-3}}~A_{*,-1}R_{\rm t, 19}^{-2}$. On the
other hand, the assumption that $4\pi R_{\rm t}^3 n_{\rm t}m_{\rm p}
\sim M_{\rm ej}$ requires that $n_{\rm t}\sim 1~{\rm
cm^{-3}}~(M_{\rm ej}/10M_\odot)R_{t,19}^{-3}$. So we have a density
contrast $n_{\rm t}/n_{\rm w}\sim 10^3$. The forward shock expanding
into the dense shell will have a pressure $\sim \beta^2 n_{\rm
t}m_{\rm p}c^2/3$, which is much higher than that of the shocked
wind medium ($\sim \beta^2 n_{\rm w}m_{\rm p}c^2/3$). A pressure
balance will be established by a strong reverse shock penetrating
into the shocked wind medium. Therefore the forward shock velocity
is much smaller than $\beta\mid_{\rm shocked~ wind~ medium}$ and can
not accelerate protons to an energy $\sim 10^{18}$ eV. The reverse
shock with a velocity $\beta_{\rm r}\approx \beta\mid_{\rm shocked~
wind~ medium}$ plausibly plays a more important role in accelerating
high energy CR protons. The shocked wind medium has only a very
small mass (relative to $M_{\rm ej}$). The reverse shock gets weak
after penetrating into the dense HN outflow which has a density
comparable to $n_{\rm t}$. Then the forward shock velocity increases
and significant CR acceleration in the forward shock front is
possible. A detailed numerical calculation, like Ptuskin \&
Zirakashvili's (2005), is needed to draw further conclusions.

\subsubsection{Gyroresonant stochastic acceleration model}\label{sec:GSAM}
The maximum energy-gain rate due to the stochastic Fermi
acceleration for marginally relativistic shock can be estimated as
(Dermer 2001b)
\begin{equation}
{dE_{_{\rm CR}} \over dR} \approx {\varepsilon_{\rm turb}(v-1)\over
2^{3/2}}ZeB_* \beta^2 ({2^{1/2}E_{\rm CR} \over Z e B_* f_\Delta R
\beta})^{v-1}, \label{eq:Dermer01}
\end{equation}
where $Z$ is the atomic number, $\varepsilon_{\rm turb}$ is the
ratio of plasma turbulence to the shock energy density, $B_*\approx
0.4~n^{1/2}\varepsilon_B^{1/2}$ Gauss, $f_\Delta \sim 1/12$ is the
ratio of the width of the swept medium by the shock to $R$ (Dermer
\& Humi 2001), and $v$ is the spectrum index of the turbulence
($v=5/3$ for Kolmogorov turbulence and $3/2$ for Kraichnan
turbulence).

Dermer (2001b) took a $\beta\sim {\rm const.}$, integrated
eq.(\ref{eq:Dermer01}) over $R$, then got $E_{\rm max}(R)$. However,
currently $\beta$ evolves with $R$. As shown below, the smaller the
radius, the larger the $\beta$ and the higher the $E_{\rm max}$.
Very energetic CRs can be accelerated at early times but can not be
accelerated continually because of the adiabatic cooling. Taking
into account the adiabatic cooling effect, eq.(\ref{eq:Dermer01})
takes the new form
\begin{equation}
{dE_{_{\rm CR}} \over dR}\approx {\varepsilon_{\rm turb}(v-1)\over
2^{3/2}}ZeB_* \beta^2 ({2^{1/2}E_{\rm CR} \over Z e B_* f_\Delta R
\beta})^{v-1}-{E_{_{\rm CR}} \over R}. \label{eq:Dermer01-b}
\end{equation}
Now $E_{\rm max}$ can be estimated by setting ${dE_{_{\rm CR}} \over
dR}=0$, then we have
\begin{equation}
E_{\rm max} \approx [{\varepsilon_{\rm turb}(v-1)\beta \over 2
f_\Delta}]^{1/(2-v)}{Z e B_* f_\Delta R \beta\over \sqrt{2}}.
\label{eq:Fan-08}
\end{equation}

\emph{\textbf{ISM medium.}} In this case,
we have
\begin{eqnarray}
&&E_{\rm max}{\rm (ISM)} \sim Z n_0^{1/2}\epsilon_{B,-1}^{1/2}R_{19} \nonumber\\
&&~~~\left\{%
\begin{array}{ll}
   10^{16}~{\rm eV}~({\varepsilon_{\rm turb}\over 0.5})^3 \beta_{-1}^4
   (12f_\Delta)^{-2} & (v=5/3), \\
   10^{17}~{\rm eV}~({\varepsilon_{\rm turb}\over 0.5})^2
    \beta_{-1}^3 (12f_\Delta)^{-1} & (v=3/2).\\
\end{array}%
\right. \label{eq:M1}
\end{eqnarray}

The energy conservation $4\pi R^3 \beta^2 n m_p c^2/3\approx
E(>\beta)$ yields $R \approx 10^{19} ~{\rm cm}~ A_{52.7}^{1/3}
\beta_{-1}^{-(q+2)/3} n_0^{-1/3}$. Combining with Eq. (\ref{eq:M1}),
we have
\begin{equation}
E_{\rm max}{\rm (ISM)} \propto \beta^{5-v-q(2-v)\over 3(2-v)},
\label{eq:M2}
\end{equation}
i.e., $E_{\rm max}{\rm (ISM)} \propto \beta^{(7-q)/3}$ for $v=3/2$ and $\propto
\beta^{(10-q)/3}$ for $v=5/3$, both are sensitive to $\beta$.

\emph{\textbf{WIND medium.}} In the termination wind shock model,
the stellar wind profile may hold up to a distance $\sim 10^{18}$ cm
(e.g., Chevalier et al. 2004). In this case,
$n=3\times10^{35}~A_*R^{-2}~{\rm cm^{-3}}$. Now $B_*\approx 0.2
A_{*,-1}^{1/2}\epsilon_{\rm B,-1}^{1/2}R_{17}^{-1}~{\rm Gauss}$ and
\begin{eqnarray}
&&E_{\rm max}{\rm (wind)} \sim Z A_{*,-1}^{1/2}\epsilon_{B,-1}^{1/2}  \nonumber\\
&&~~~\left\{%
\begin{array}{ll}
   2\times 10^{15}~{\rm eV}~({\varepsilon_{\rm turb}\over 0.5})^3 \beta_{-1}^4
   (12f_\Delta)^{-2} & (v=5/3), \\
    2\times 10^{16}~{\rm eV}~({\varepsilon_{\rm turb}\over 0.5})^2
    \beta_{-1}^3 (12f_\Delta)^{-1} & (v=3/2).\\
\end{array}%
\right. \label{eq:3}
\end{eqnarray}\\

As shown in Eqs. (\ref{eq:M1}) and (\ref{eq:3}), for
$\varepsilon_{\rm turb}\sim 0.5$ and $v=(3/2,~5/3)$, at $\beta \sim
\beta_o \sim 0.1$, we have $E_{\rm max}\sim (10^{17},~10^{16})Z$ eV.
Below we focus on the case of $v=3/2$, because in the case of
$v=5/3$ the request of $E_{\rm max}(\beta_o) \sim 3\times 10^{17}$
eV is more difficult to satisfy. For $\beta\sim 0.5$, the stochastic
gyroresonant acceleration is able to accelerate protons to $\sim
10^{19}$ eV  (see also Dermer 2001a). The accelerated particle
spectrum is thus ($v=3/2$)
\begin{equation}
{dN_{\rm CR} \over dE_{\rm CR}} \propto
\left\{%
\begin{array}{ll}
   E_{\rm CR}^{-(2+\delta)-{3q\over
7-q}} & {\rm for~} k=0, \\
    E_{\rm CR}^{-(2+\delta)-{q\over
3}} & {\rm for~} k=2.\\
\end{array}%
\right. \label{eq:R1}
\end{equation}
As shown in section 2.1, the main deceleration of the HN outflow is
very likely to be in an homogenous medium. The accelerated protons
have a spectrum $dN/dE_{_{\rm CR}}\propto E_{\rm
CR}^{-(2.4+3q/(7-q))}$. To match the observation $\Delta \gamma({\rm
II})\approx -3q/(7-q)\sim -0.9$, we need
\[
q\sim 1.6.
\]
This is surprisingly close to the value $\sim 1.7$ that is inferred
from the optical modeling of SN 2003lw. Detailed optical modeling of
more GRB-associated HN explosions is highly needed to better
constrain $q$ and then confirm or rule out our interpretation.

If GRB-associated HNe expand into a Wind bubble-like medium, a
flatter CR spectrum in the higher energy range would appear. At a
small radius (say, $<10^{18}$ cm), the medium is free Wind-like and
the accelerated particle spectrum is $\propto E_{\rm
CR}^{-(2+\delta)-q/3}$, which then gets steepened by a factor of
$q(2+q)/[3(7-q)]\sim 0.4$ for $q\sim 1.6$ after entering the
ISM-like medium. Such a flattening seems not enough to match the
observation $\Delta \gamma ({\rm III}) \sim 0.6$. So CRs above the
ankle may be mainly from AGNs, as indicated by the recent analysis
of the correlation of the highest-energy CRs with nearby
extragalactic objects by the Pierre Auger Collaboration
\cite{PAC07}.

The rate of local GRB-associated HNe only accounts for $\sim
(0.1-0.5)\%$ of that for all local SNe \cite{DV06,sod07}. The
typical energy of these HNe, however, is tens times larger than that
of the normal SNe. Roughly, we expect that a fraction $\sim 10\%$ of
CR protons at 3 PeV could be attributed to GRB-associated HNe. It is
enough to match the observation \cite{ulr04,Horand07}. So the CR
proton spectrum in the energy range of $10^{16}-10^{18.5}$ eV may be
quantitatively interpreted.

\section{Discussion and Summary}
The particle acceleration in marginally relativistic HN shocks are
discussed. The GRB-associated HN outflows are assumed to have a
fairly steep energy distribution against their velocities, i.e.,
$E(\geq \beta)\propto \beta^{-q}$ for $q\sim 1.7$, as inferred from
the optical modelling of SN 2003lw (see section 2.1 for details). A
significant fraction of a HN's kinetic energy is carried by the
material moving with a velocity $>\beta_o (\sim 0.1)$, driving an
energetic shock wave into the surrounding medium. The cosmic ray
protons above the second knee but below the ankle may be accelerated
by the HN shocks in the velocity range of $\beta \sim (1-4)\beta_o$.
To satisfy this velocity bound, the HN outflows associated with GRBs
must have reached a very large radius where the surrounding medium
is very likely to be ISM-like (see section 2.2 for details). With
this argument, the self-magnetized shock acceleration model adopted
in Wang et al. (2007) would yield a very steep spectrum that is
inconsistent with the observation unless the medium surrounding the
HN is a free wind holding up to a radius $R_{\rm dec}\sim 10~{\rm
kpc}~(M_{\rm ej}/10M_\odot)A_{*,-1}^{-1}$. Such a request seems
difficult to satisfy. A highly speculative solution is that the
particle acceleration mainly occurs in a narrow ``dense" shell that
terminates the free wind at a radius $\sim 10^{19}$ cm (see the last
paragraph of section \ref{sec:SMSAM} for details).

In this work, we find that for $q\sim 1.6$, the stochastic
gyroresonant acceleration model can account for the spectrum change
of high energy protons around the second knee (see section
\ref{sec:GSAM} for details). As a consequence, the stochastic
gyroresonant acceleration mechanism in relativistic GRB forward
shock may account for part of the ultra-high energy CRs ($\sim
10^{20}$ eV), as suggested in Dermer (2001b, 2007) and Dermer \&
Humi (2001). A typical
\[q\approx {-7\Delta \gamma({\rm II})\over 3-\Delta \gamma({\rm
II})} \sim 1.6~~{\rm for}~\Delta \gamma({\rm II}) \sim -0.9,\] if
confirmed in future optical modelling of the GRB-associated HN
explosions, will be a crucial evidence for our current speculation.

\section*{Acknowledgments}
We thank an anonymous referee and Charles D. Dermer for constructive
comments, and Jin-Song Deng, Zhuo Li and Dong Xu for
communication/discussion. This work is supported by a (postdoctoral)
grant from the Danish National Science Foundation, the National
Natural Science Foundation (grant 10673034) of China and a special
grant of Chinese Academy of Sciences.

\end{document}